\documentclass[aps,pre,nofootinbib,showpacs,tightenlines,preprint,titlepage,amsmath]{revtex4}
\usepackage{bm}

\newcommand{\be}{\begin{equation}}
\newcommand{\ee}{\end{equation}}
\newcommand{\bea}{\begin{eqnarray}}
\newcommand{\eea}{\end{eqnarray}}
\newcommand{\cC}{\ensuremath{\mathcal{C}}}
\newcommand{\cP}{\ensuremath{\mathcal{P}}}
\newcommand{\cT}{\ensuremath{\mathcal{T}}}

\newcommand{\half}{\mbox{$\textstyle{\frac{1}{2}}$}}

\begin{document}
\title{$\cP\cT$-Symmetric Versus Hermitian Formulations of Quantum Mechanics}

\author{Carl M. Bender}\email{cmb@wustl.edu}
\author{Jun-Hua Chen}\email{jchena@artsci.wustl.edu}
\author{Kimball A. Milton}\email{milton@nhn.ou.edu}\altaffiliation{Permanent
address: Oklahoma Center for High Energy Physics and Homer L. Dodge Department
of Physics and Astronomy, University of Oklahoma, Norman, OK 73019, USA}

\affiliation{Department of Physics, Washington University, St. Louis, MO 63130,
USA}

\date{\today}

\begin{abstract} 
A non-Hermitian Hamiltonian that has an unbroken $\cP\cT$ symmetry can be
converted by means of a similarity transformation to a physically equivalent
Hermitian Hamiltonian. This raises the following question: In which form of the
quantum theory, the non-Hermitian or the Hermitian one, is it easier to perform
calculations? This paper compares both forms of a non-Hermitian $ix^3$
quantum-mechanical Hamiltonian and demonstrates that it is much harder to
perform calculations in the Hermitian theory because the perturbation series for
the Hermitian Hamiltonian is constructed from divergent Feynman graphs. For the
Hermitian version of the theory, dimensional continuation is used to regulate
the divergent graphs that contribute to the ground-state energy and the
one-point Green's function. The results that are obtained are identical to those
found much more simply and without divergences in the non-Hermitian
$\cP\cT$-symmetric Hamiltonian. The $\mathcal{O}(g^4)$ contribution to
the ground-state energy of the Hermitian version of the theory involves graphs
with overlapping divergences, and these graphs are extremely difficult to
regulate. In contrast, the graphs for the non-Hermitian version of the theory
are finite to all orders and they are very easy to evaluate.
\end{abstract}

\maketitle
\section{Introduction}
\label{s1}
In 1998 it was shown using perturbative and numerical arguments that the
non-Hermitian Hamiltonians
\be
H=p^2+x^2(ix)^\epsilon\quad(\epsilon\geq0)
\label{e1}
\ee
have real positive spectra \cite{r1,r2}. It was argued in these papers that
the reality of the spectrum was due to the unbroken $\cP\cT$ symmetry of the
Hamiltonians. A rigorous proof of reality was given by Dorey {\it et al.}
\cite{r3}.

Later, in 2002 it was shown that the Hamiltonian in (\ref{e1}) describes unitary
time evolution \cite{r4}. In Ref.~\cite{r4} it was demonstrated that it is
possible to construct a new operator called $\cC$ that commutes with the
Hamiltonian $H$. It was shown that the Hilbert space inner product with respect
to the $\cC\cP\cT$ adjoint has a positive norm and that the time evolution
operator $e^{iHt}$ is unitary. Evidently, Dirac Hermiticity of the Hamiltonian
is not a necessary requirement of a quantum theory; unbroken $\cP\cT$ symmetry
is sufficient to guarantee that the spectrum of $H$ is real and positive and
that the time evolution is unitary. (In this paper we indicate that a
Hamiltonian is Hermitian in the Dirac sense by writing $H=H^\dag$, where the
symbol $\dag$ indicates Dirac Hermitian conjugation, that is, the combined
operations of complex conjugation and matrix transposition: $H^\dag\equiv
H^{*{\rm T}}$.)

A recipe for constructing $\cC$ was given in Ref.~\cite{r5}. The procedure is
to solve the three simultaneous algebraic equations satisfied by $\cC$:
\be
\cC^2=1,\quad [\cC,\cP\cT]=0,\quad [\cC,H]=0.
\label{e2}
\ee
The recipe in Ref.~\cite{r5} has been used to find the $\cC$ operator for
various quantum field theories \cite{r6,r7,r8}. This recipe produces the $\cC$
operator as a product of the exponential of an antisymmetric Hermitian operator
$Q$ and the parity operator $\cP$:
\be
\cC=e^Q\cP.
\label{e3}
\ee
As an example, we consider the $\cP\cT$-symmetric non-Hermitian Hamiltonian
\be
H=\half p^2+\half x^2+ix.
\label{e4}
\ee
For this Hamiltonian, the exact $Q$ operator is given by
\be
Q=-2p.
\label{e5}
\ee

A natural question to ask is whether there is a Hamiltonian that is Hermitian in
the Dirac sense and is equivalent to a non-Hermitian $\cP\cT$-symmetric
Hamiltonian $H$. Mostafazadeh has shown that there is a Hermitian operator
$\rho$ that may to used to perform a similarity transformation on $H$,
\be
h=\rho^{-1}H\rho,
\label{e6}
\ee
to produce a new Hamiltonian $h$ that is Hermitian in the Dirac sense \cite{r9}.
The operator $\rho$ is just the square-root of the (positive) $\cC\cP$ operator:
\be
\rho=e^{Q/2}.
\label{e7}
\ee
The Hamiltonian $h$ that results from the similarity transformation (\ref{e6})
has been studied perturbatively by Jones \cite{r10} and Mostafazadeh \cite{r11}.

We summarize briefly the work in Refs.~\cite{r10,r11}. One can verify that the
Hamiltonian $h$ produced by the similarity transformation (\ref{e6}) is
Hermitian by taking the Hermitian conjugate of $h$:
\be
h^\dag=\left( e^{-Q/2}He^{Q/2}\right)^\dag=e^{Q/2}H^\dag e^{-Q/2}.
\label{e8}
\ee
Next, one uses the $\cP\cT$ symmetry of $H$ to replace $H^\dag$ by $\cP H\cP$,
\be
h^\dag=e^{Q/2}\cP H\cP e^{-Q/2},
\label{e9}
\ee
and one uses the identity (\ref{e3}) to rewrite (\ref{e9}) as
\be
h^\dag=e^{-Q/2}\cC H\cC e^{Q/2}.
\label{e10}
\ee
But $\cC$ commutes with $H$, so
\be
h^\dag=e^{-Q/2}He^{Q/2}=h,
\label{e11}
\ee
which establishes the Hermiticity of $h$.

We illustrate this transformation by using the Hamiltonian (\ref{e4}).
The similarity transformation (\ref{e11}) using (\ref{e5}) gives
\be
h=\half p^2+\half x^2+\half,
\label{e12}
\ee
which is clearly Hermitian.

To see that $H$ and $h$ have the same spectra, one can multiply the eigenvalue
equation for $H$, $H\Phi_n=E_n\Phi_n$, on the left by $e^{-Q/2}$:
\be
e^{-Q/2}He^{Q/2}e^{-Q/2}\Phi_n=E_ne^{-Q/2}\Phi_n.
\label{e13}
\ee
Thus, the eigenvalue problem for $h$ reads $h\phi_n=E_n\phi_n$, where the
eigenvectors $\phi_n$ are given by $\phi_n\equiv e^{-Q/2}\Phi_n$. More
generally, the association between states $|A\rangle$ in the Hilbert space for
the $\cP\cT$-symmetric theory and states $|a\rangle$ in the Hilbert space for
the Hermitian theory is given by
\be
|a\rangle=e^{-Q/2}|A\rangle.
\label{e14}
\ee

The Hermitian theory whose dynamics is specified by $h$ has the standard Dirac
inner product:
\be
\langle a|b\rangle\equiv (|a\rangle)^\dag\cdot|b\rangle.
\label{e15}
\ee
However, the inner product for the non-Hermitian theory whose dynamics is
governed by $H$ is the $\cC\cP\cT$ inner product explained in Ref.~\cite{r4}:
\be
\langle A|B\rangle_{\cC\cP\cT}\equiv\left(\cC\cP\cT|A\rangle\right)^{\rm T}\cdot
|B\rangle.
\label{e16}
\ee
If $|a\rangle$ and $|b\rangle$ are related to $|A\rangle$ and $|B\rangle$ by
$|a\rangle=e^{-Q/2}|A\rangle$ and $|b\rangle=e^{-Q/2}|B\rangle$ according to
(\ref{e14}), then the two inner products in (\ref{e15}) and (\ref{e16}) are
identical. To show this one can argue as follows:
\bea
&&\langle a|=(|a\rangle)^\dag=(|a\rangle)^{\ast{\rm T}}=(\cT|a\rangle)^{\rm T}=
\left(\cT e^{-Q/2}|A\rangle\right)^{\rm T}=\left(e^{Q/2}\cT|A\rangle\right)^{\rm
T}\nonumber\\
&&\qquad\qquad=\left(e^{-Q/2}e^Q\cP\cP\cT|A\rangle\right)^{\rm T}=\left(e^{-Q/2}
\cC\cP\cT|A\rangle\right)^{\rm T}=(\cC\cP\cT|A\rangle)^{\rm T}e^{Q/2}.
\label{e17}
\eea
Thus, $\langle a|b\rangle=\langle A|B\rangle_{\cC\cP\cT}$.

In this paper we discuss the Hermitian Hamiltonian corresponding to the cubic
non-Hermitian $\cP\cT$-symmetric Hamiltonian
\be
H=\half p^2+\half x^2+ig x^3.
\label{e18}
\ee
This is the quantum-mechanical analog of the field-theoretic Hamiltonian density
\be
{\cal H}=\half(\partial\varphi)^2+\half m^2\varphi^2+ig\varphi^3,
\label{e19}
\ee
which is a non-Hermitian scalar quantum field theory that has appeared often in
the literature. This quantum field theory describes the Lee-Yang edge
singularity \cite{r12} and arises in Reggeon field theory \cite{r13}. The
construction given in Ref.~\cite{r5} of the $\cC$ operator for this quantum
field theory demonstrates that this model is a physical unitary quantum theory
and not an unrealistic mathematical curiosity.

The question to be addressed in this paper is whether the Hermitian form of
the Hamiltonian (\ref{e18}) is more useful or less useful than the non-Hermitian
form. To answer this question, in Sec.~\ref{s2} we calculate the ground-state
energy to order $g^2$ using Feynman graphical methods for both the Hermitian and
the non-Hermitian versions of the theory. We focus on graphical methods here
because only graphical methods can be used as a perturbative approach in quantum
field theory. We find that for the non-Hermitian version of the theory the
Feynman rules are simple and the calculation is utterly straightforward. In
contrast, for the Hermitian version of the theory the Feynman rules are
significantly more complicated and lead to divergent integrals that must be
regulated. In Sec.~\ref{s3} we show how to calculate the one-point Green's
function in both versions of the theory to order $g^3$. Again, we encounter
divergent graphs in the Hermitian theory, and these graphs must be regulated to
obtain the correct answer. In Sec.~\ref{s4} we show that the Feynman rules in
the Hermitian theory become increasingly complicated as one goes to higher
orders in perturbation theory. One is inevitably led to very difficult divergent
integrals that involve overlapping divergences. In contrast, the calculation for
the $\cP\cT$-symmetric version of the theory is extremely simple and only
contains finite graphs. We conclude in Sec.~\ref{s5} that the Hermitian version
of the $\cP\cT$-symmetric theory is impractical.

\section{Calculation of the Ground-State Energy to Order $g^2$}
\label{s2}

The Schr\"odinger eigenvalue problem corresponding to the quantum-mechanical
Hamiltonian (\ref{e18}) is easy to solve perturbatively, and we can calculate
the ground-state energy as a series in powers of $g^2$. The fourth-order result
is
\be
E_0=\half+{\textstyle\frac{11}{8}}g^2-{\textstyle\frac{465}{32}}g^4+\mathcal{O}
(g^6).
\label{e20}
\ee
However, our ultimate objective is to study $\cP\cT$-symmetric quantum field
theories, and therefore we need to construct Feynman-diagrammatic methods to
solve for the Green's functions of the theory.

\begin{figure}[b!]
\vspace{2.1in}
\includegraphics{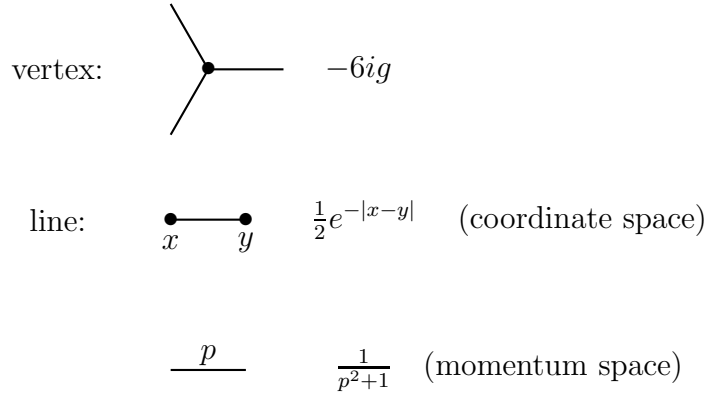}
\caption{Feynman rules for the Lagrangian (\ref{e23}). For this simple local
trilinear interaction the Feynman graphs are built from three-point vertices
connected with lines. The line amplitudes in both coordinate space and momentum
space are shown.}
\label{f1}
\end{figure}

For any quantum field theory the perturbation expansion of the ground-state
energy is the negative sum of the connected Feynman graphs having no external
lines. To evaluate Feynman graphs we must first determine the Feynman rules,
which are obtained from the Lagrangian. Thus, we begin by constructing the
Lagrangian corresponding to the Hamiltonian $H$ in (\ref{e18}):
\be
L=\half(p\dot x +\dot x p)-H.
\label{e21}
\ee
Because the interaction term is local (it depends only on $x$ and not on $p$),
the formula for $\dot x$ is simple:
\be
\dot x=p.
\label{e22}
\ee
Thus, we have
\be
L=\half\dot x^2-\half x^2-igx^3.
\label{e23}
\ee
From (\ref{e23}) we read off the Euclidean Feynman rules: The three-point vertex
amplitude is $-6ig$. In coordinate space a line connecting vertices at $x$ and
$y$ is represented by $\half e^{-|x-y|}$ and in momentum space the line
amplitude is $\frac{1}{p^2+1}$. These Feynman rules are illustrated in
Fig.~\ref{f1}.

\begin{figure}[b!]
\vspace{1.4in}
\includegraphics{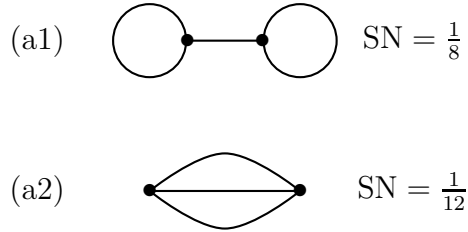}
\caption{The two connected vacuum graphs, labeled (a1) and (a2), contributing to
the ground-state energy of $H$ in (\ref{e18}) to order $g^2$. The symmetry
numbers for each graph are indicated.}
\label{f2}
\end{figure}

In order $g^2$ there are two connected graphs that contribute to the
ground-state energy, and these are shown in Fig.~\ref{f2}. The symmetry number
for graph (a1) is $\frac{1}{8}$ and the symmetry number for graph (a2) is $\frac
{1}{12}$. Both graphs have vertex factors of $-36g^2$. The evaluation of the
Feynman integrals for (a1) and (a2) gives $V/4$ and $V/12$, respectively, where
$V=\int dx$ is the volume of coordinate space. Thus, the sum of the graph
amplitudes is $-\frac{11}{8}g^2V$. The contribution to the ground-state energy
is the negative of this amplitude divided by $V$: $E_2=\frac{11}{8}g^2$, which
easily reproduces the $g^2$ term in (\ref{e20}).

We showed in Sec.~\ref{s1} that the energy levels of the Hermitian Hamiltonian
$h$ that is obtained by means of the similarity transformation (\ref{e6}) are
identical to those of $H$. Our objective here is to recalculate the $g^2$ term
in the expansion of the ground-state energy in (\ref{e20}) using the Feynman
rules obtained from the transformed Hamiltonian $h$. The first step in this
calculation is to construct the operator $Q$, which is given in Ref.~\cite{r5}
as
\be
Q=\left(-{\textstyle\frac{4}{3}}p^3-2S_{1,2}\right)g+
\left({\textstyle\frac{128}{15}}p^5+{\textstyle\frac{40}{3}}S_{3,2}+8S_{1,4}-12p
\right)g^3+{\mathcal O}(g^5),
\label{e24}
\ee
where the symbol $S_{m,n}$ represents a totally symmetric combination of $m$
factors of $p$ and $n$ factors of $x$.

One can use (\ref{e6}) -- (\ref{e7}) to construct $h$. The result given in
Refs.~\cite{r10,r11} is
\bea
h&=&\half p^2+\half x^2+\left({\textstyle\frac{3}{2}}x^4+3S_{2,2}-\half\right)
g^2\nonumber\\
&&\mbox{}+\left(-{\textstyle\frac{7}{2}}x^6-{\textstyle\frac{51}{2}}S_{2,4}
-36S_{4,2}+2p^6+{\textstyle\frac{15}{2}}x^2+27p^2\right)g^4+\mathcal{O}(g^6).
\label{e25}
\eea
In order to obtain the Feynman rules we must now construct the corresponding
Hermitian Lagrangian $\ell$. To do so, we must replace the operator $p$ with
the operator $\dot x$ by using the formula
\be
p=\dot x-6g^2s_{1,2},
\label{e26}
\ee
where $s_{m,n}$ represents a totally symmetric combination of $m$ factors of
$\dot x$ and $n$ factors of $x$. The result for the Hermitian Lagrangian $\ell$
is
\bea
\ell&=&\half \dot x^2-\half x^2-\left({\textstyle\frac{3}{2}}x^4+3s_{2,2}-\half
\right)g^2\nonumber\\
&&\mbox{}+\left({\textstyle\frac{7}{2}}x^6+{\textstyle\frac{87}{2}}s_{2,4}
+36s_{4,2}-2\dot x^6-{\textstyle\frac{27}{2}}x^2-27\dot x^2\right)g^4+\mathcal{
O}(g^6).
\label{e27}
\eea

\begin{figure}[b!]
\vspace{2.6in}
\includegraphics{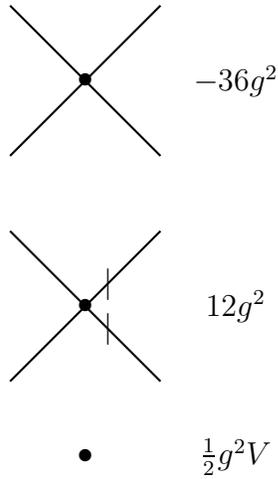}
\caption{The three Euclidean-space vertices to order $g^2$ for the Hermitian
Lagrangian $\ell$ in (\ref{e27}). Note that the second vertex has tick marks
on two of the legs. These tick marks indicate coordinate-space derivatives that
arise because of derivative coupling. Derivative coupling results in divergent
Feynman graphs.}
\label{f3}
\end{figure}

\begin{figure}[b!]
\vspace{2.2in}
\includegraphics{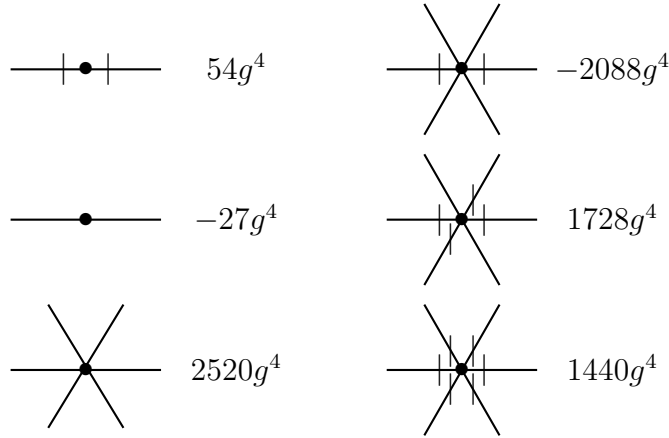}
\caption{The six Euclidean-space vertices to order $g^4$ for the Hermitian
Lagrangian $\ell$ in (\ref{e27}). Note that four of the vertices have tick marks
on the legs. These tick marks indicate derivative coupling.}
\label{f4}
\end{figure}

From this Lagrangian we can read off the Euclidean-space Feynman rules. Unlike
the $\cP\cT$ version of the theory, increasingly many new vertices appear in
every order of perturbation theory. The three vertices to order $g^2$ are shown 
in Fig.~\ref{f3} and the six vertices to order $g^4$ are shown in Fig.~\ref{f4}.
Note that some of the lines emerging from the vertices have tick marks. A tick
mark indicates a derivative in coordinate space and a factor of $ip$ in
momentum space. The tick marks are a result of the {\it derivative} coupling
terms in the Lagrangian $\ell$. As we will see, the derivative coupling
gives rise to divergent Feynman graphs.

\begin{figure}[b!]
\vspace{2.0in}
\includegraphics{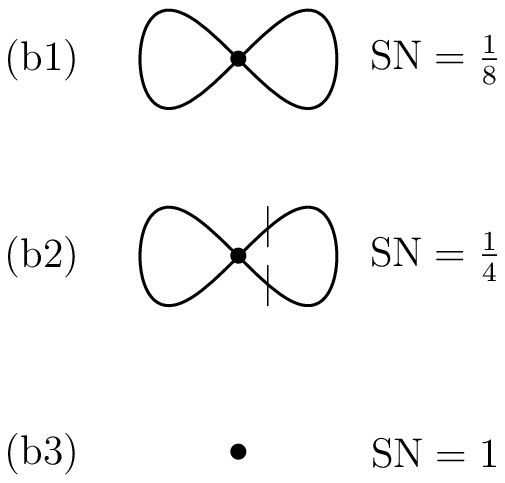}
\caption{The three graphs contributing to the ground-state energy of the
Hermitian Lagrangian $\ell$ in (\ref{e27}) in order $g^2$. Note that while
graphs (b1) and (b3) are finite, the Feynman integral for graph (b2) diverges
and must be regulated to obtain a finite result.}
\label{f5}
\end{figure}

We now use the Feynman rules in Fig.~\ref{f3} to construct the vacuum graphs
contributing to the ground-state energy in order $g^2$. These graphs are
shown in Fig.~\ref{f5}. The simplest of these three graphs is (b3) because
there is no Feynman integral to perform. This graph arises from the constant
term in $\ell$ in (\ref{e27}). The value of this graph is simply $\half g^2V$.

Graph (b1) has symmetry number $\frac{1}{8}$ and vertex factor $-36g^2$ and the
Feynman integral in momentum space is
\be
\left(\int_{-\infty}^\infty\frac{dp}{2\pi}\,\frac{1}{p^2+1}\right)^2=\frac{1}
{4}.
\label{e28}
\ee
The integrals associated with this graph are convergent. The value of graph
(b1) is therefore $-\frac{9}{8}g^2V$, where the factor of $V$ comes from the
translation invariance of the graph.

The interesting graph is (b2). The symmetry number is $\frac{1}{4}$, the vertex
factor is $12g^2$, and the Feynman integral in momentum space is
\be
\int_{-\infty}^\infty\frac{dp}{2\pi}\,\frac{p^2}{p^2+1}
\int_{-\infty}^\infty\frac{dq}{2\pi}\,\frac{1}{q^2+1}.
\label{e29}
\ee
The $q$ integral is convergent and gives the value $\half$. However, the $p$
integral is divergent. We therefore regulate it using dimensional continuation
and represent its value as the limit as the number of dimensions approaches 1:
\be
\lim_{D\to1}2\int_0^\infty\frac{r^{D-1}dr}{2\pi}\,\frac{r^2}{r^2+1}=\lim_{D\to1}
\frac{\Gamma\left(1+\half D\right)\Gamma\left(-\half D\right)}{2\pi}=-\frac{1}
{2}.
\label{e30}
\ee
Hence, the value of graph (b2) is $-\frac{3}{4}g^2V$, where again the volume
factor $V$ comes from translation invariance. Adding the three graphs (b1),
(b2), and (b3), dividing by $V$, and changing the sign gives the result $\frac{1
1}{8}g^2$, which reproduces the result in (\ref{e20}). This is a more difficult
calculation than that using the Feynman rules in Fig.~\ref{f1} because we
encounter a divergent graph. It is most surprising to find a divergent graph in
one-dimensional quantum field theory (quantum mechanics). The infinite graph
here is not associated with a renormalization of a physical parameter in the
Lagrangian. Rather, it is an artifact of the derivative coupling terms that
inevitably arise from the similarity transformation (\ref{e6}).

Here is a simple model that illustrates the use of dimensional continuation as
a means of regulating Feynman graphs: Consider the quadratic Lagrangian
\be
L=\half\dot x^2-\half x^2-\half g\dot x^2.
\label{eee1}
\ee
The corresponding Hamiltonian is 
\be
H=\half p^2+\half x^2+\frac{g}{2-2g}p^2.
\label{eee2}
\ee
The ground-state energy $E_0$ for $H$ in (\ref{eee2}) is 
\be
E_0=\half(1-g)^{-1/2}.
\label{eee3}
\ee
The Euclidean Feynman rules for $L$ in (\ref{eee1}) are elementary. The
amplitude for a line is given in Fig.~\ref{f1} and there is a two-tick
two-point vertex with amplitude $g$. (This vertex has the form of the first
vertex shown in Fig.~\ref{f4}.) The graphs contributing to the ground-state
energy are all polygons. The $n$th-order graph has $n$ vertices; its symmetry
number is $\frac{1}{2n}$ and its vertex amplitude is $g^n$. The total graphical
contribution to the ground-state energy is
\be
E_0-\half=-\sum_{n=1}^\infty\frac{g^n}{2n}\int_{-\infty}^\infty\frac{dp}{2\pi}
\,\frac{p^{2n}}{(p^2+1)^n}.
\label{eee4}
\ee
Each of the integrals in (\ref{eee4}) is divergent, but we regulate the
integrals as in (\ref{e30}):
\be
\lim_{D\to1}2\int_0^\infty\frac{r^{D-1}dr}{2\pi}\,\left(\frac{r^2}{r^2+1}
\right)^n=\lim_{D\to1}\frac{\Gamma\left(n+\half D\right)\Gamma\left(-\half D
\right)}{2\pi(n-1)!}=-\frac{\Gamma\left(n+\half\right)}{\pi^{1/2}(n-1)!}.
\label{eee5}
\ee
Therefore, (\ref{eee4}) becomes
\be
E_0=\half\sum_{n=0}^\infty g^n\frac{\Gamma\left(n+\half\right)}{\pi^{1/2}n!}
=\half(1-g)^{-1/2},
\label{eee6}
\ee
which verifies the result in (\ref{eee3}).

This dimensional-continuation procedure is effective because it extracts the
correct finite contribution from each of the divergent graphs. However, this
procedure is much more difficult to apply when there are graphs having
overlapping divergences, as we shall see in Sec.~\ref{s4}.

\section{Calculation of the One-Point Green's Function}
\label{s3}

The connection in (\ref{e14}) between states in the Hermitian and the
non-Hermitian $\cP\cT$-symmetric theories implies the following relation between
an operator $O$ in the non-Hermitian $\cP\cT$-symmetric theory and the
corresponding operator $\tilde O$ in the Hermitian theory:
\be
\tilde O=e^{-Q/2}Oe^{Q/2}.
\label{e31}
\ee
Using this connection, we now calculate the one-point Green's function $G_1$ in
both versions of the theory to order $g^3$. Again, we find that the calculation
in the non-Hermitian theory is extremely simple, but that in the Hermitian
theory the calculation again involves divergent graphs that must be regulated.

\begin{figure}[b!]
\vspace{2.9in}
\includegraphics{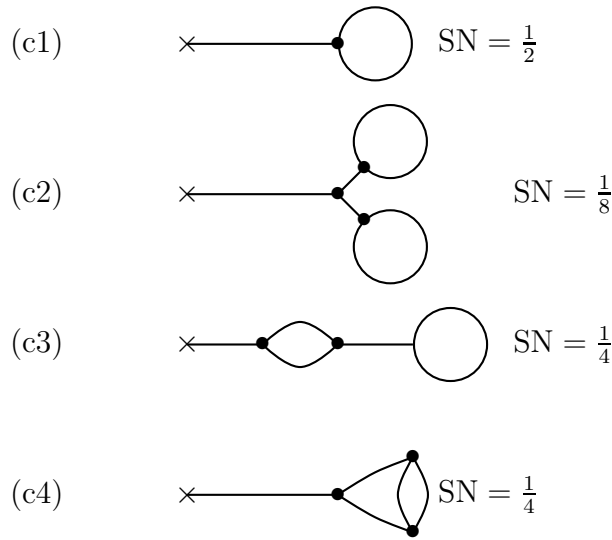}
\caption{The Feynman graphs contributing to the one-point Green's function $G_1$
of the non-Hermitian Hamiltonian $H$ through order $g^3$. Graph (c1) is of order
$g$ and graphs (c2) -- (c4) are of order $g^3$.}
\label{f6}
\end{figure}

The graphs contributing to $G_1=\langle0|x|0\rangle_{\cC\cP\cT}$ through order
$g^3$ in the non-Hermitian theory defined by $H$ in (\ref{e18}) or,
equivalently, $L$ in (\ref{e23}) are shown in Fig.~\ref{f6}. Each of these
graphs is finite and is easily evaluated. The result is that
\be
G_1=-{\textstyle\frac{3}{2}}ig+{\textstyle\frac{33}{2}}ig^3+\mathcal{O}(g^5).
\label{e32}
\ee

Next, we calculate the identical one-point Green's function in the Hermitian
theory. To do so, we need to transform the field $x$ to the corresponding field
$\tilde x$ in the Hermitian theory using (\ref{e31}):
\bea
\tilde x&=&e^{-Q/2}xe^{Q/2}\nonumber\\
&=&x-i\left(x^2+2p^2\right)g+\left(2S_{2,1}-x^3\right)g^2+i\left(20p^4+24S_{2,2}
+5x^4-6\right)g^3+\mathcal{O}(g^4)\nonumber\\
&=&x-i\left(x^2+2\dot x^2\right)g+\left(2s_{2,1}-x^3\right)g^2+i\left(20\dot
x^4+48s_{2,2}+5x^4-6\right)g^3+\mathcal{O}(g^4),
\label{e33}
\eea
where we have replaced $p$ in favor of $\dot x$ using (\ref{e26}). (This result
may be found in Refs.~\cite{r10,r11} to order $g^2$.)

\begin{figure}[b!]
\vspace{1.1in}
\includegraphics{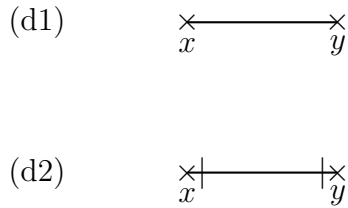}
\caption{The two graphs contributing to the one-point Green's function $G_1$ in
the Hermitian theory to order $g$. Note that graph (d1) is finite, while graph
(d2) diverges and must be regulated to give a finite result.}
\label{f7}
\end{figure}

Using (\ref{e33}) we can construct the graphs contributing to $G_1$ to order
$g$ (see Fig.~\ref{f7}). Graph (d1) is finite and has the value $-\half ig$.
However, graph (d2) is infinite and must be regulated using dimensional
continuation. The result is $-ig$. Combining these two graphs, we obtain the
term of order $g$ in (\ref{e32}).

\begin{figure}[b!]
\vspace{2.1in}
\includegraphics{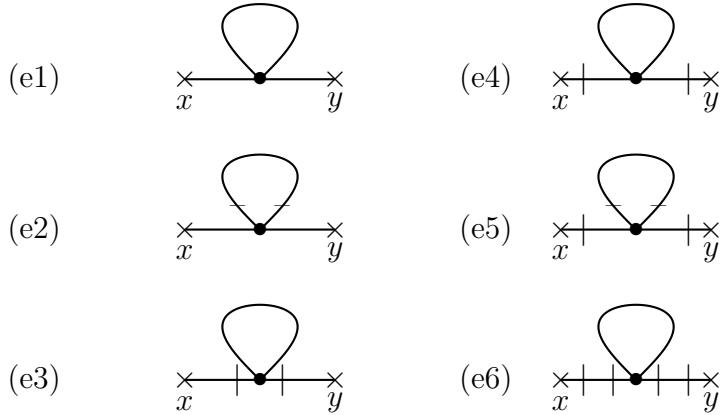}
\caption{The six connected graphs contributing to the one-point Green's function
$G_1$ for the Hermitian theory to order $g^3$. Graphs (e1), (e3), and (e4) are
finite, but the remaining graphs are divergent and must be regulated.}
\label{f8}
\end{figure}

The calculation of $G_1$ to order $g^3$ in the Hermitian theory is much more
complicated. First, we must construct the six connected graphs arising from the
expectation value of $\tilde x$ in (\ref{e33}) (see Fig.~\ref{f8}). Three of
these graphs, (e1), (e3), and (e4), are finite. The remaining graphs are
divergent and must be regulated using dimensional continuation. There are four
more disconnected graphs arising from the expectation values of the terms $20
\dot x^4$, $48s_{2,2}$, $5x^4$, and $-6$. Two of these graphs must also be
regulated. Finally, combining the contributions of these ten graphs, we
successfully reproduce the $\mathcal{O}(g^3)$ result $\frac{33}{2}ig^3$ in
(\ref{e32}). We emphasize that the calculation for the Hermitian theory is
orders of magnitude more difficult than the corresponding calculation for the
non-Hermitian $\cP\cT$-symmetric theory.

\section{Higher-order Calculation of the Ground-State Energy}
\label{s4}

In this section we extend the calculation of the ground-state energy that is
described in Sec.~\ref{s2} to next order in powers of $g^2$. We will see that
this calculation is completely straightforward in the non-Hermitian theory,
while it is nearly impossible in the Hermitian theory. We show that the
difficulty is not just due to the arithmetic difficulty of sorting through large
numbers of graphs, but rather is one of principle. The problem is that we
encounter two graphs with overlapping divergences, and calculating the numerical
values of the corresponding regulated graphs remains an unsolved problem, even
in one-dimensional field theory (quantum mechanics)!

\begin{figure}[b!]
\vspace{3.9in}
\includegraphics{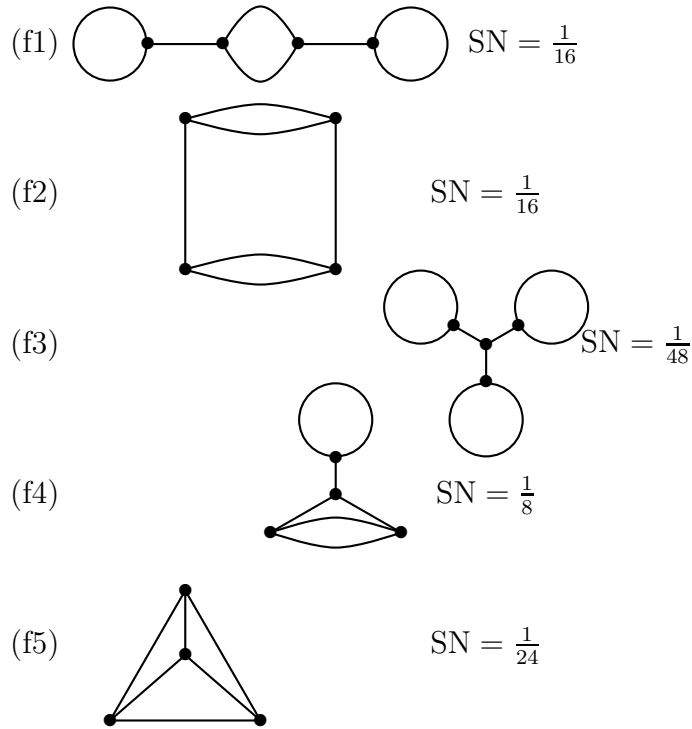}
\caption{The five vacuum graphs contributing to the ground-state energy of the
non-Hermitian $\cP\cT$-symmetric Hamiltonian $H$ in (\ref{e18}) to order $g^4$.
These graphs are all finite and very easy to evaluate.}
\label{f9}
\end{figure}

There are five graphs (f1) -- (f5) contributing in order $g^4$ to the
ground-state energy of the non-Hermitian Hamiltonian $H$ in (\ref{e18}). These
are shown in Fig.~\ref{f9}. The symmetry numbers for these graphs are indicated
in the figure. The vertex factors for all these graphs are $1296g^4$. The
Feynman integrals for these graphs are $\frac{1}{16}V$ for (f1), $\frac{11}{864}
V$ for (f2), $\frac{1}{8}V$ for (f3), $\frac{1}{36}V$ for (f4), and $\frac{1}{96
}V$ for (f5). Thus, the sum of the graphs is $\frac{465}{32}V$. The negative of
this amplitude divided by $V$ is $E_4=-\frac{465}{32}g^4$. This reproduces the
order $g^4$ term in the perturbation expansion for the ground-state energy in
(\ref{e20}).

\begin{figure}[b!]
\vspace{6.8in}
\includegraphics{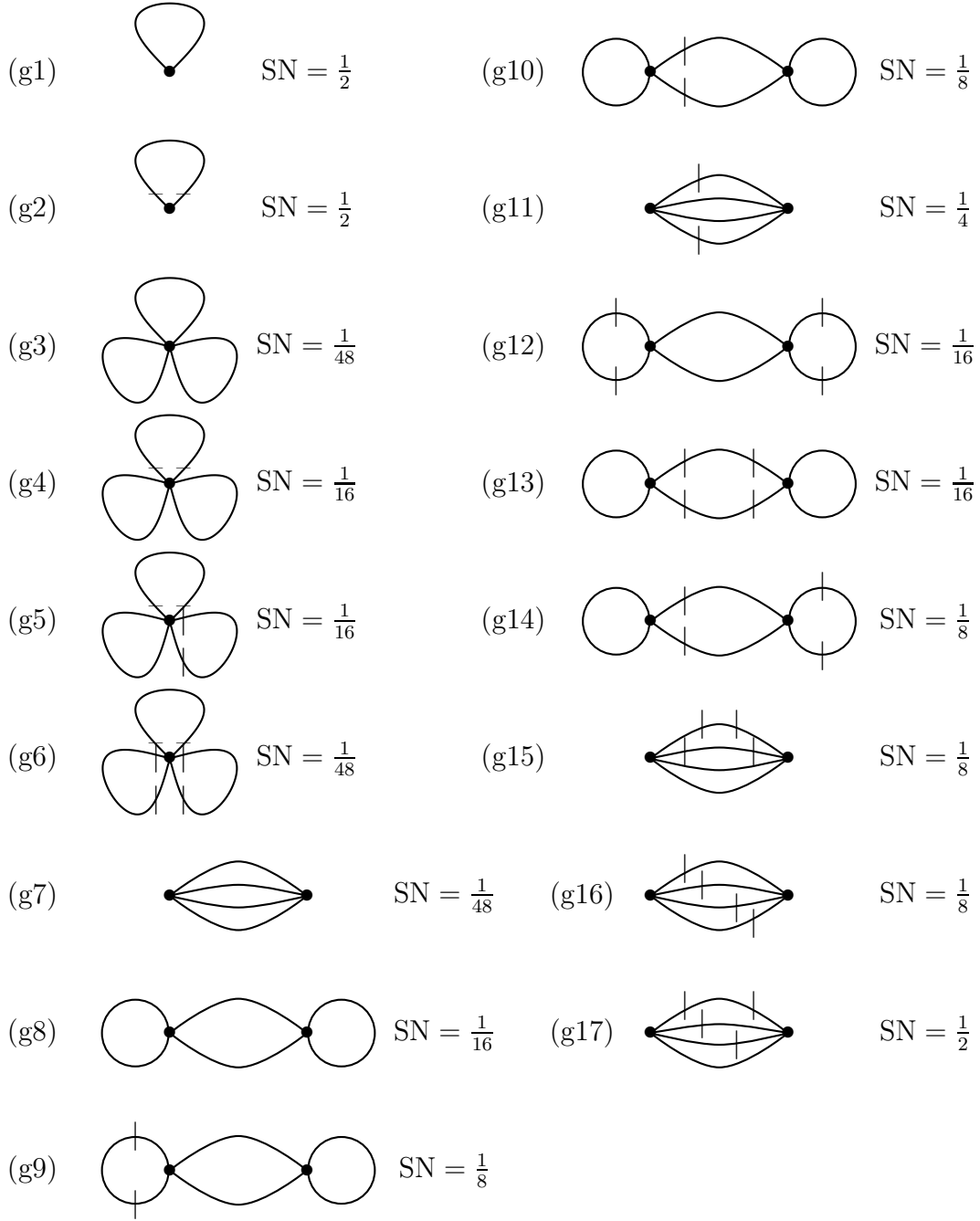}
\caption{The seventeen graphs contributing to the order $g^4$ term in the
perturbation expansion for the ground-state energy of the Hermitian Hamiltonian
$h$ in (\ref{e25}). Note that ten of these graphs have divergent Feynman
integrals. Of these ten, eight are relatively easy to regulate using dimensional
continuation. However, graphs (g15) and (g17) have overlapping divergences, and
are therefore extremely hard to evaluate.}
\label{f10}
\end{figure}

There are seventeen graphs of order $g^4$ contributing to the ground-state
energy of the Hermitian Hamiltonian (\ref{e25}). These graphs, along with their
symmetry numbers, are shown in Fig.~\ref{f10}. Seven of these graphs, (g1),
(g3), (g7), (g8), (g10), (g11), and (g16) are finite and easy to calculate.
The Feynman integrals for the remaining graphs are all infinite and must be
regulated. Dimensional continuation can be readily implemented as in
(\ref{e30}) except for the graphs (g15) and (g17). These two graphs are
extremely difficult to regulate because they have overlapping divergences.
It is most dismaying to find Feynman graphs having overlapping divergences in
one-dimensional quantum field theory! Since the $g^4$ contribution to the
ground-state energy is given in (\ref{e20}), we can deduce that the sum of the
regulated values of these two graphs (multiplied by their respective symmetry
numbers and vertex factors) must be $\frac{21}{16}g^4V$. However, we are unable
to find a simple way to obtain this result.

\section{Concluding Remarks}
\label{s5}

This study was motivated by the concern that the mechanics of solving problems
in quantum field theory might not work in non-Hermitian theories. The usual
techniques rely on the use of the Schwinger action principle, the construction
of functional integrals, and the identification of and application of Feynman
rules. These procedures are conventionally formulated in a Hermitian setting.
The surprise is that all of these standard techniques work perfectly in a
non-Hermitian context, but that they are much too difficult to apply if the
non-Hermitian theory is first transformed to the equivalent Hermitian one.

We conclude by citing the comment of Jones in the second paper in
Ref.~\cite{r10} regarding the critique of $\cP\cT$-symmetric theories in
Ref.~\cite{r14}. Jones writes, ``Clearly, this [Eq.~(\ref{e25})] is not a
Hamiltonian that one would have contemplated in its own regard were it not
derived from [Eq.~(\ref{e18})]. It is for this reason that we disagree with the
contention of Mostafazadeh \cite{r14} that, `A consistent probabilistic
$\cP\cT$-symmetric quantum theory is doomed to reduce to ordinary quantum
mechanics.'\," Mostafazadeh appears to be correct in arguing that a
$\cP\cT$-symmetric theory can be transformed to a Hermitian theory by means of a
similarity transformation. However, we have demonstrated that the difficulties
with the Hermitian theory are severe and virtually insurmountable because this
theory possesses a Feynman perturbation expansion that becomes increasingly
divergent as one goes to higher order. The divergences are not removable by
renormalization, but rather are due to increasingly singular derivative
interactions. In contrast, the non-Hermitian $\cP\cT$-symmetric theory is
completely free from all such difficulties.

\acknowledgments{KAM is grateful for the hospitality of the Physics Department
at Washington University. We thank the U.S. Department of Energy for partial
financial support of this work.}

\end{document}